\documentclass[
journal=jacsat, 
manuscript=article]{achemso}
\setkeys{acs}{articletitle = false}

\usepackage[version=3]{mhchem} 
\usepackage{xcolor,soul} 

\author{Bodo Zibrowius}
\email{bodo.zibrowius@posteo.de}
\affiliation
{45468 M\"ulheim an der Ruhr, Germany}

\author{Michael Fischer}
\email{michael.fischer@uni-bremen.de}
\affiliation
{Crystallography \& Geomaterials, Faculty of Geosciences, University of Bremen, Klagenfurter Stra{\ss}e
2-4, 28359 Bremen, Germany}
\alsoaffiliation{Bremen Center for Computational Materials Science, University of Bremen, 28359 Bremen, Germany}
\alsoaffiliation{MAPEX Center for Materials and Processes, University of Bremen, 28359 Bremen, Germany}

\title{$^{27}$Al NMR spectroscopic and DFT computational study of the quadrupole coupling of aluminium in two polymorphs of the complex aluminium hydride \ce{CsAlH4}}

\begin{document}

\begin{abstract}
The quadrupole coupling constant $C_{\text{Q}}$ and the asymmetry parameter $\eta$ of the aluminium nuclei in two polymorphs of the complex aluminium hydride \ce{CsAlH4} are determined from both $^{27}$Al MAS~NMR spectra and $^{27}$Al~NMR spectra recorded for stationary samples by using the Solomon echo sequence. The accuracy with which these parameters can be determined from the  static spectra (\ce{CsAlH4(\textit{o})}: $C_{\text{Q}}=(1.42\pm0.01)$\,MHz, $\eta=(0.62\pm0.01)$ and \ce{CsAlH4(\textit{t})}: $C_{\text{Q}}=(1.43\pm0.02)$\,MHz, $\eta<0.03$) seems to be slightly higher than via the MAS approach.

The experimentally determined parameters ($\delta_{\text{iso}}$, $C_{\text{Q}}$ and $\eta$) are compared with those obtained from DFT-GIPAW (density functional theory - gauge-including projected augmented wave) calculations. 
When using DFT-optimized structures, the magnitude of the quadrupole coupling constant is overestimated by about 45\% for both polymorphs. Further calculations in which the geometry of the \ce{AlH4} tetrahedra was varied show a high sensitivity of $C_{\text{Q}}$
on the H--Al--H angles in particular. Modest changes in the angles on the order of one to three degrees are sufficient to achieve near-perfect agreement between GIPAW calculations and experiment. The deviations found for the DFT-optimized structures are explained with the neglect of thermal motion, which typically leads to a reduction of distortions of the \ce{AlH4} tetrahedra.
From a broader perspective, the uncertainty in the positions of the hydrogen atoms renders the accurate reproduction or prediction of quadrupole coupling constants for aluminium hydrides challenging.

\end{abstract}

\newpage

\section{Introduction}

Because of their potential application as hydrogen storage materials, complex aluminium hydrides have been studied rather extensively  in the past two decades. Following the seminal paper by  Bogdanovi\'{c} and Schwickardi \cite{Bogdanovic97}, the reversible dehydrogenation of \ce{NaAlH4} under the influence of catalysts was the target of many experimental and theoretical studies. \cite{Schueth04, Orimo07, Bogdanovic09, Frankcombe12, Li13, Milanese18, Zhao21} Although complex aluminium hydrides with heavier cations are obviously less promising candidates for a reversible hydrogen storage, they have been included into these studies, mainly to get more insights into the mechanisms of dehydrogenation and re-hydrogenation. In the course of these investigations, several new complex aluminium hydrides have been discovered and structurally characterized. \cite{Suarez19} 

About a decade ago, it was demonstrated that caesium aluminium tetrahydride (\ce{CsAlH4}) can easily be synthesized by ball-milling stoichiometric amounts of \ce{NaAlH4} and \ce{CsCl}. \cite{Krech14} However, the product was found to be a mixture of a previously described orthorhombic phase (\ce{CsAlH4(\textit{o})}) and a hitherto unknown tetragonal phase (\ce{CsAlH4(\textit{t})}). The crystal structures of both polymorphs were solved from X-ray powder data. \cite{Krech14}  In the same paper, it was shown that these polymorphs can reversibly be transformed into each other by thermal or mechanical treatment without any observable degradation. Furthermore, the polymorphs were found to be easily distinguishable by $^{27}$Al and $^{133}$Cs magic-angle spinning (MAS) NMR spectroscopy. A subsequent neutron powder diffraction study of the corresponding deuterides,\cite{Bernert15} that allowed the hydrogen positions to be determined, confirmed that \ce{CsAlH4(\textit{o})} crystallizes in space group \textit{Pnma}. For \ce{CsAlH4(\textit{t})}, the authors postulated a disordered \textit{I}$4_{1}/$\textit{a} structure with hydrogen atoms on two distinct Wyckoff 16$f$ positions with an occupancy of 0.5 each.

The first aim of the present paper is to demonstrate that $^{27}$Al NMR spectroscopy can provide a much more comprehensive characterisation of the \ce{CsAlH4} polymorphs. In particular, we want to determine the parameters of the quadrupole coupling of the aluminium nuclei: the quadrupole coupling constant $C_{\text{Q}}$, which is proportional to both the strength of the electric field gradient (EFG) and the electric quadrupole moment of the nucleus under study, and the asymmetry parameter $\eta$.\cite{FreudeHaase, quadnmr, Man11} In NMR spectroscopy, quadrupole interaction is often regarded as drawback because of its detrimental effect on the spectral resolution. However, the parameters $C_{\text{Q}}$ and $\eta$ are directly linked to the local geometry around the nucleus under study. For example, if the nucleus is located on a symmetry axis $C_{\text{n}}$ with $n\geq3$, the EFG is axially symmetric and $\eta$ is zero. In a perfect cubic environment, there is no field gradient and, consequently, no quadrupole interaction. Hence, any quadrupolar nucleus can in principle be used to probe the local structure in a solid material. 

On the other hand, modern DFT methods allow the parameters of the chemical shift and the quadrupole coupling to be calculated on the basis of the structural data.\cite{Zwanziger12}  In our previous paper \cite{ZF24} on \ce{NaAlH4} and \ce{KAlH4}, we found that the degree of agreement between quadrupole coupling data derived from NMR spectra and those from DFT calculations based on the available structure data from powder diffraction was markedly different for the two aluminium hydrides  studied. While an excellent agreement between the experimental and the calculated quadrupole coupling constant was found for \ce{NaAlH4}, the strength of the quadrupole coupling for \ce{KAlH4} was overestimated in the DFT calculations by about 30\%. Therefore, the second aim of present paper is to better understand the factors influencing the reliability of the calculated data. The caesium aluminium hydrides \ce{CsAlH4(\textit{t})} and \ce{CsAlH4(\textit{o})} are structurally closely related to \ce{NaAlH4} and \ce{KAlH4}, respectively.

\section{Experimental}

\subsection{Synthesis}

As previously described in more detail, \cite{Krech14} \ce{CsAlH4} can easily be synthesized by ball milling \ce{NaAlH4} with \ce{CsCl}. Dissolution of the material prepared in this salt metathesis reaction in diglyme and subsequent precipitation yields in general a mixture of two polymorphs of \ce{CsAlH4}. Almost phase-pure \ce{CsAlH4(\textit{o})} can be obtained when the percipitate is heated up to 200\,$^{\circ}$C and kept at this temperature for 2\,h.   Almost phase-pure \ce{CsAlH4(\textit{t})} can be obtained when the percipitate or \ce{CsAlH4(\textit{o})} is ball milled under a pressure  of 200\,bar \ce{H2}. The samples used in the present paper were prepared according to this post-synthesis procedure. They are identical to the samples S4 (\ce{CsAlH4(\textit{o})}) and S5 (\ce{CsAlH4(\textit{t})}) of the earlier paper.\cite{Krech14}
	
	All syntheses and operations were performed under argon using dried and oxygen-free solvents. The MAS rotors were filled and capped in a glove box and transferred to the spectrometer in argon-filled vials.

\subsection{Solid-state NMR spectroscopy}

The $^{27}$Al NMR spectra were recorded on a Bruker Avance\,III\,HD 500WB spectrometer using double-bearing MAS probes (DVT BL4) at a resonance frequency of 130.3\,MHz. The chemical shift was referenced relative to an external 1.0\,M aqueous solution of aluminium nitrate. The same solution was used for determining the flip-angle. 

For the $^{27}$Al MAS~NMR spectra, single $\pi$/12 pulses ($t_{\text{p}}$ = 0.6\,$\mu$s) were applied with a repetition time of 2\,s (4,000~scans) at several spinning frequencies ($\nu_{\text{MAS}}$) between 4 and 10\,kHz.  High-power proton decoupling (SPINAL-64\cite{Fung00}) was used for all $^{27}$Al~NMR spectra shown in this paper. The magic angle was adjusted by maximizing the rotational echoes of the $^{23}$Na resonance of solid \ce{NaNO3}. 

The $^{27}$Al NMR spectra of stationary samples were acquired using the Solomon echo sequence with two pulses of the same length $t_{\text{p}}$ separated by a delay $\tau$.\cite{Solomon58} The theoretical background of the Solomon echo technique was outlined in more detail in our previous paper.\cite{ZF24} For $^{27}$Al ($I = \tfrac {5}{2}$), this pulse sequence generates echoes at $\tau$/2, $\tau$, 3$\tau$/2, 2$\tau$, and 3$\tau$ after the second pulse.\cite{Man97, Man00}  The echo at $\tau$, which is used here to obtain the spectra, contains the spectral information of both satellite transitions. Which of the various echoes are experimentally observed can to a certain extent be influenced by the phase cycling applied.\cite{Bonhomme04} A 16-step phase cycle that was originally proposed by Kunwar et al.\cite{Kunwar86} was used for the acquisition of the Solomon echo data shown here. This phase cycle is known to effectively cancel spurious signals from the NMR probe.\cite{Man92}  Two strong rf pulses ($\nu_{\text{rf}}$ $\approx$ 100\,kHz) with a length $t_{\text{p}}$ = 0.9\,$\mu$s were applied. To minimize the interfering effect of the echo generated at 2$\tau$ after the second pulse, a pulse spacing $\tau$ of 0.6\,ms was used. 24,000 scans were accumulated with a repetition time of 2\,s. To start Fourier transform at the top of the echo at $\tau$, a pre-acquisition delay slightly shorter than $\tau$ and an appropriate number of left shifts were applied (dwell time: 0.1\,$\mu$s). 

For observing NMR lines that have a total width of about 1\,MHz, the asymmetric response function of the probe circuit has to be taken into account, both for MAS and static spectra. \cite{Giavani02,Giavani04,Groszewicz17} We tried to reduce the intensity and phase distortions that can be caused by the response function by choosing an appropriate length of the cable between pre-amplifier and probe and by tuning the probe slightly off-resonance. The symmetry of the response function was checked with a well crystalline \ce{NaAlH4} sample that exhibits extremely sharp maxima of the satellite transitions.\cite{ZF24}

The spectra simulations were performed using the solids lineshape analysis module implemented in the TopSpin\texttrademark\ 3.2 NMR software package from Bruker BioSpin GmbH.

\subsection{DFT-GIPAW Calculations}
\label{calculations}

DFT structure optimizations and calculations of NMR parameters were carried out with CASTEP.\cite{Clark2005,Milman2010} In keeping with our previous study,\cite{ZF24} the exchange-correlation functional devised by Perdew, Burke, and Ernzerhof (PBE) was used.\cite{Perdew1996}. The calculations employed on-the-fly generated ultrasoft pseudopotentials, and the plane wave cutoff energy was set to 800\,eV. The first Brillouin zone of \ce{CsAlH4}(\textit{o}) was sampled using a $4\times6\times5$ Monkhorst-Pack mesh of $k$-points, corresponding to 18 irreducible points. A $7\times7\times3$ $k$-mesh (26 irreducible points) was used for \ce{CsAlH4}(\textit{t}). In the structure optimizations, cell parameters were fixed to experimental values (see below), relaxing all atomic positions while obeying symmetry constraints. Using a BFGS optimizer, the following convergence criteria were enforced: maximal residual force on an atom below 0.001\,eV/\AA, maximal atomic displacement from one step to the next below 0.0005\,\AA.

Starting structures of \ce{CsAlH4}(\textit{o}) and \ce{CsAlH4}(\textit{t}) were taken from the literature. For the orthorhombic phase, initial atomic coordinates were taken from the isotypic crystal structure of \ce{KAlH4}, using cell parameters from the X-ray diffraction study of Krech et al. ($a$ = 9.8973\,\AA, $b$ = 6.1507\,\AA, $c$ = 7.9197\,\AA).\cite{Krech14} For the tetragonal phase, atomic coordinates from the crystal structure of \ce{NaAlH4} were combined with cell parameters determined by Krech et al. ($a$ = 5.6732\,\AA, $c$ = 14.2795\,\AA). Essentially identical total energies and NMR parameters were obtained when using input coordinates from the neutron diffraction study by Bernert et al.\cite{Bernert15} These calculations used the coordinates from column 2 of Table 1 of that work for \ce{CsAlH4}(\textit{o}) and those contained in column 2 of Table 2 for \ce{CsAlH4}(\textit{t}). For the tetragonal phase, separate optimizations assuming full occupancy of either of the two deuterium positions, refined in the work of Bernert et al.\cite{Bernert15} with an occupancy of 0.5, converged to the same final structure. Although optimizations of \ce{CsAlH4}(\textit{t}) started in space group $I4_1/a$, postulated in that work, they always resulted in a rotation of the \ce{AlH4} tetrahedra that increased the space group symmetry to $I4_1/a m d$ in the final relaxed structure. In fact, a symmetry search on the DFT-optimized structure obtained previously by Bernert et al.\cite{Bernert15} (coordinates taken from column 5 of their Table 2), reportedly having space group symmetry $I4_1/a$, revealed that this structure actually obeys $I4_1/a m d$ symmetry, with the deuterium atoms lying on Wyckoff position 16\textit{h} ($y=\frac{3}{4}$). As already pointed out in the same work, an earlier computational study proposing \ce{CsAlH4}(\textit{t}) did not explicitly state the space group symmetry, but labelled the H position as Wyckoff position 16\textit{h}, which is consistent with $I4_1/a m d$ symmetry as there is no such Wyckoff position in $I4_1/a$.\cite{Ravindran06} Altogether, our DFT calculations did not provide any evidence for a stable local minimum in space group $I4_1/a$, and point to $I4_1/a m d$ as the symmetry of \ce{CsAlH4}(\textit{t}). 

DFT calculations of the isotropic shielding parameter $\sigma_{\text{iso}}$, the quadrupole coupling constant $C_{\text{Q}}$, and the asymmetry parameter $\eta$ also used the CASTEP code, which permits the calculation of NMR parameters in the framework of the gauge-including projector augmented wave (GIPAW) method.\cite{Pickard2001,Profeta2003,Yates2007,Bonhomme2012} The calculations were performed on structures with fully optimized atomic coordinates, as described above, and for derived structures in which the geometry of the \ce{AlH4} tetrahedra was varied (see below). Settings of the DFT-GIPAW calculations (exchange-correlation functional, pseudopotentials, cutoff energy, $k$-mesh size) were identical to those presented above for the structure optimizations. For \textsuperscript{27}Al, isotropic shielding parameters $\sigma_{\text{iso,DFT}}$ were converted into chemical shifts $\delta_{\text{iso,DFT}}$ using the linear relationship reported in our previous work,\cite{ZF24} which was established on the basis of DFT calculations using an analogous setup for six alkali aluminum hydrides: 
\begin{equation}
  \delta_{\text{iso,DFT}} = 514.3\,\text{ppm} - 0.9064 \cdot \sigma_{\text{iso,DFT}}\text{.}
   \label{eq:delta}
\end{equation}

In order to evaluate the influence of changes in the geometry of the \ce{AlH4} tetrahedra on the NMR parameters, additional DFT-GIPAW calculations were carried out for structures in which Al--H distances and H--Al--H angles were varied across a range of relevant values. The respective variations are described in detail in the corresponding part of the Results and Discussion section.

While not being in the focus of the present study, isotropic \textsuperscript{133}Cs NMR chemical shifts were also obtained from the DFT-GIPAW calculations. For \textsuperscript{133}Cs, the following equation was used for the conversion of isotropic shieldings  $\sigma_{\text{iso,DFT}}$ into isotropic chemical shifts $\delta_{\text{iso,DFT}}$: 

\begin{equation}
  \delta_{\text{iso,DFT}} = 5447.8\,\text{ppm} + 218.5 \,\text{ppm} - \sigma_{\text{iso,DFT}}\text{.}
   \label{eq:delta_Cs}
\end{equation}
Here, 5447.8\,ppm is the $\sigma_{\text{iso,DFT}}$ value obtained for crystalline CsCl from DFT-GIPAW calculations with analogous settings as described above, and 218.5\,ppm corresponds to the difference in the chemical shift between solid CsCl and a 1.0 M aqueous solution of CsCl.\cite{Hayashi90}

\section{Results and Discussion}

\subsection{NMR}

\begin{figure}
  \includegraphics[width=8cm]{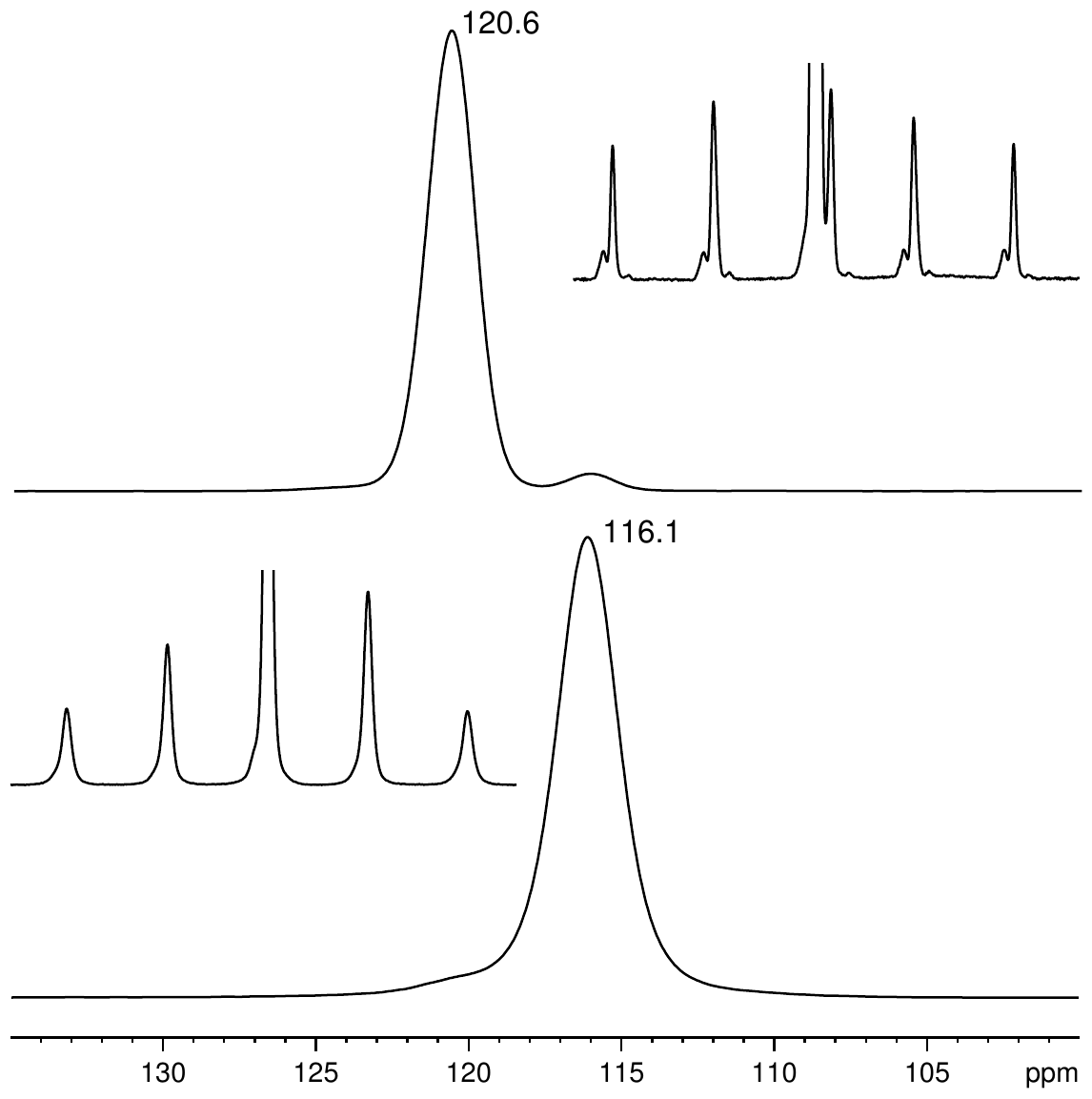}
  \caption{ $^{27}$Al MAS~NMR spectra  of the two polymorphs of \ce{CsAlH4}. The spectrum at the top was recorded for a sample that contains mainly \ce{CsAlH4(\textit{o})} ($\geq$\,96\%)  and the one at the bottom for a sample that contains mainly  \ce{CsAlH4(\textit{t})} ($\geq$\,98\%). Both spectra were taken at a spinning speed $\nu_{\text{MAS}}=4$\,kHz.  The full width at half height (FWHH) of the resonance line is 240\,Hz and 310\,Hz for the orthorhombic and the tetragonal sample, respectively. A Lorentzian line broadening with $LB= 5$\,Hz was applied. The insets show the first- and second-order spinning sidebands in addition to the centrebands. Here, the centrebands are cut off at about 4\% and 25\% of their maximum intensity for the orthorhombic and tetragonal sample, respectively.}
  \label{fgr:MAS_CT}  
\end{figure}

Figure~\ref{fgr:MAS_CT} shows the $^{27}$Al~MAS~NMR spectra of two almost phase-pure \ce{CsAlH4} samples that were prepared according to the post-synthesis procedure described in the Experimental section.  Precipitates obtained directly after the synthesis are usually mixtures of these two polymorphs.\cite{Krech14} Both resonance lines are in the chemical shift range usually observed for complex aluminium hydrides that contain only isolated AlH$_{4}^{-}$ ions.\cite{Tarasov08} With full widths at half height (FWHH) that vary between 240\,Hz (1.8\,ppm)  and 360\,Hz (2.7\,ppm) from sample to sample, the difference in the line positions of about 4.3\,ppm is more than sufficient for a reliable quantification of as-synthesized \ce{CsAlH4} by deconvolution of the $^{27}$Al~MAS~NMR spectra if the centrebands and first-order spinning sidebands are taken into account. The lines observed for the tetragonal phase are usually slightly broader than those for the orthorhombic phase.  Although the two samples used in the present study contain small amounts of the respective other phase, they will be referred to as \ce{CsAlH4(\textit{o})} and \ce{CsAlH4(\textit{t})} in the following. The percentages given in the caption of Figure~\ref{fgr:MAS_CT} are conservative estimates based on spectra deconvolution. For the sample referred to as  \ce{CsAlH4(\textit{t})}, the amount of \ce{CsAlH4(\textit{o})} is probably overestimated because of the overlap of its resonance line with the centreband of the satellite transitions of the main phase. 

The centrebands of the central transition shown in Figure~\ref{fgr:MAS_CT} are rather symmetric and show no indication of second-order quadrupole interaction.\cite{FreudeHaase, quadnmr} As observed earlier for the aluminium tetrahydrides \ce{KAlH4} \cite{ZF19} and \ce{NaAlH4}\cite{ZF24}, higher spinning speeds, in particular those above 6\,kHz, result in broader lines. At a spinning speed of 10\,kHz and experimental conditions otherwise identical to those used for recording the spectra shown in Figure~\ref{fgr:MAS_CT}, we observed centrebands with FWHH of 260\,Hz and 340\,Hz for the same orthorhombic and tetragonal sample, respectively.  We assume that this broadening at higher spinning rates is caused by frictional heating, \cite{Brus00, Antonijevic05} that leads to a significant increase of the temperature in certain parts of the sample. The change in temperature is likely to  influence the chemical shift and/or the quadrupole coupling parameters. 

A closer inspection of the MAS~NMR spectra in Figure~\ref{fgr:MAS_CT} shows that the intense centrebands shown  are surrounded by wide manifolds of spinning sidebands mainly stemming from the satellite transitions.  Both the inner ($(+\tfrac{1}{2}\leftrightarrow+\tfrac{3}{2})$ and $(-\tfrac{3}{2}\leftrightarrow-\tfrac{1}{2})$) and the outer ($(+\tfrac{3}{2}\leftrightarrow+\tfrac{5}{2})$ and  $(-\tfrac{5}{2}\leftrightarrow-\tfrac{3}{2})$) satellite transitions are subjected to first-order quadrupole interaction.\cite{Man00} The insets in Figure~\ref{fgr:MAS_CT} depict only the first- and second-order spinning sidebands on both sides of the centrebands. For the orthorhombic phase the sidebands of the inner and outer satellite transitions are clearly separated from each other, whereas no such resolution is achieved for the tetragonal phase. Furthermore, there is obviously a significant contribution from the central transition to the low-order spinning sidebands for the \ce{CsAlH4(\textit{t})} sample. Both observations can be explained by the large number of crystal defects in \ce{CsAlH4(\textit{t})}, as this phase is an immediate product of a ball milling process. The rather broad reflections 
observed in the X-ray powder diffraction pattern of the tetragonal sample \cite{Krech14} support this interpretation.
 
\begin{figure}
  \includegraphics[width=8cm]{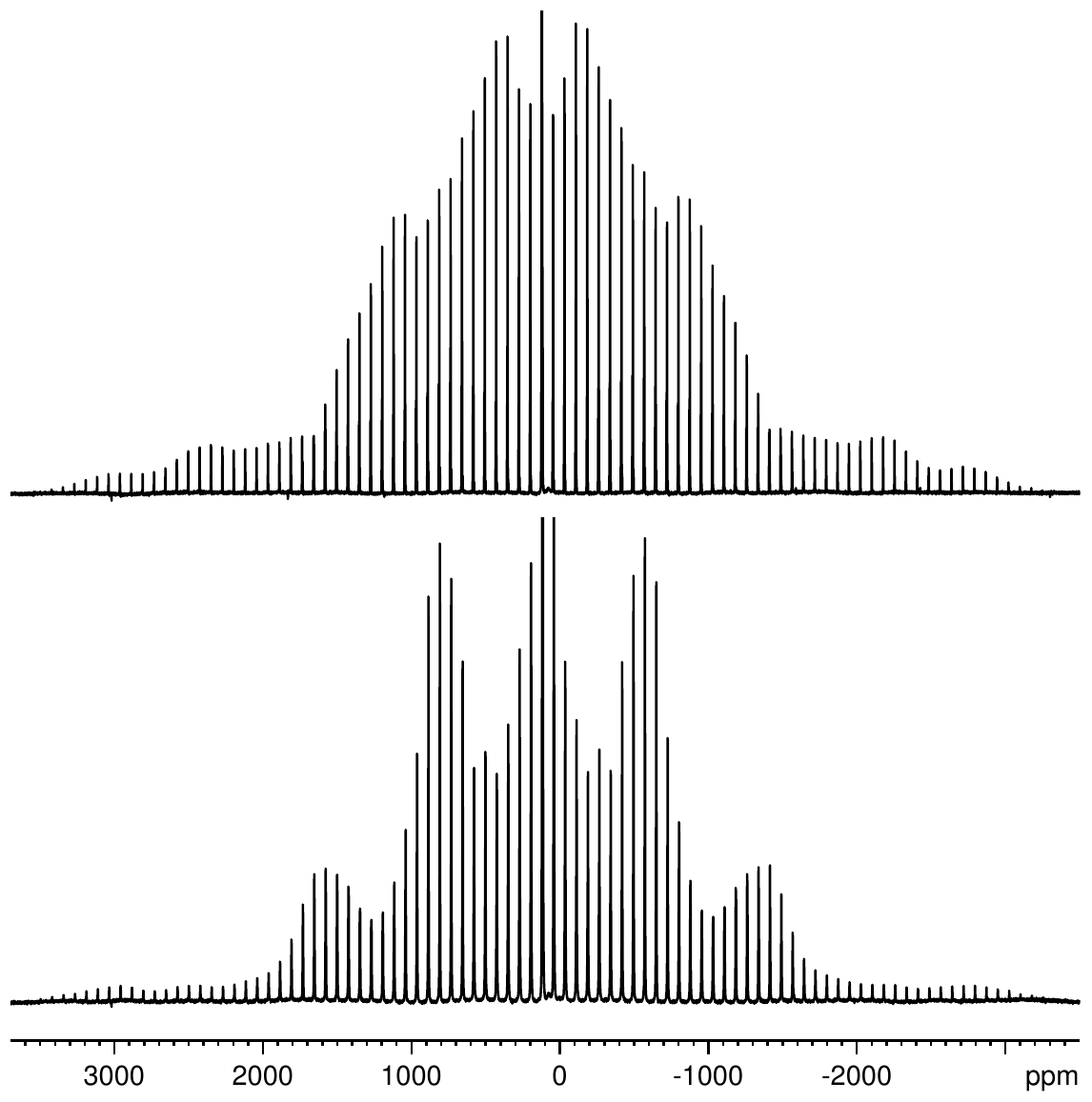}
  \caption{$^{27}$Al MAS~NMR spectra of \ce{CsAlH4(\textit{o})} (top) and  \ce{CsAlH4(\textit{t})} (bottom) measured at a spinning speed $\nu_{\text{MAS}}=10$\,kHz. These spectra show the characteristic patterns of the spinning sidebands of the satellite transitions caused by first-order quadrupole interaction. The spectra are cut off at about 10\% of the maximum intensity of the centrebands.}
  \label{fgr:MAS}  
\end{figure} 
 
The whole spinning-sideband manifolds observed for \ce{CsAlH4(\textit{o})} and \ce{CsAlH4(\textit{t})} are shown in Figure~\ref{fgr:MAS}. These patterns are strikingly different. Nevertheless, the spinning sidebands of the outer satellite transition ($m=\tfrac {5}{2}$) spread for both polymorphs over the same range, from about 3420\,ppm to about $-3180$\,ppm, i.e.\ over about 860\,kHz. The total spread of this transition, $\Delta\nu_{\text{TS}}(\tfrac {5}{2})$, is directly related to the quadrupole coupling constant $C_{\text{Q}}$:\cite{Taylor75,Freude00,ZF19}  
\begin{equation}
  C_{\text{Q}}=\frac{5}{3}\Delta\nu_{\text{TS}}(\tfrac {5}{2}).
   \label{eq:a}
\end{equation}
Hence, $C_{\text{Q}}\approx1,43$\,MHz is a good estimate for both polymorphs. It should be noted that from the NMR measurements reported here, only the magnitude of the quadrupole coupling constant $C_{\text{Q}}$ can be determined, but not the sign. The resonance lines of every pair of satellite transitions are mirror images of each other. In accordance with most of the NMR literature, \cite{quadnmr} we omit the absolute value bars for $C_{\text{Q}}$ throughout the paper. However, we give the signs of the quadrupole coupling constants obtained from DFT calculations.

At sufficiently low spinning speeds, the intensity envelope of the spinning-sideband manifold is close to the lineshape one would obtain without sample spinning. Although this condition is certainly not fulfilled for the spectra shown in Figure~\ref{fgr:MAS}, the approximate positions of the maxima of the inner satellite transition ($m=\tfrac {3}{2}$) can easily be read off these spectra. From the ratio between the splitting of the maxima  $\Delta\nu_{\text{M}}(m)$ for any satellite transition to the total spread of the spectrum $\Delta\nu_{\text{TS}}(m)$ of this transition, a fairly good estimate for the asymmetry parameter $\eta$ can be obtained via the following relation:\cite{ZF19}

\begin{equation}
  \eta=1-\frac{2\Delta\nu_M(m)}{\Delta\nu_{TS}(m)}.
   \label{eq:b}
\end{equation}

For the orthorhombic phase, the splitting between the maxima of the inner satellite transition amounts to at least 460\,ppm corresponding to $\Delta\nu_{\text{M}}(\tfrac {3}{2})= 60$\,kHz. Since the total spread of the inner satellite transition $\Delta\nu_{TS}(\tfrac {3}{2})$ is half the total spread of the outer transition $\Delta\nu_{TS}(\tfrac {5}{2})$ determined above, we obtain $\eta\approx0.7$ as an upper limit for the asymmetry parameter. For the tetragonal phase, the maxima of the inner satellite transition are separated by at least 1540\,pm corresponding to $\Delta\nu_{\text{M}}(\tfrac {3}{2})= 200$\,kHz, i.\,e. this splitting is about a quarter of $\Delta\nu_{TS}(\tfrac {5}{2})$. Hence, as expected from the crystal structure with the aluminium nucleus on a fourfold axis, there is no indication for a deviation from axial symmetry in the experimental spectrum ($\eta\approx0.$).
 
These estimates for  $C_{\text{Q}}$ and $\eta$ are rather good starting values for the simulation of the spinning-sideband manifolds. The characteristic intensity modulations of the spinning sidebands can be used to determine the parameters of the quadrupole interaction with high accuracy.\cite{Jakobsen89, Skibsted91} Recently, we have shown that the parameters of the quadrupole interaction of the aluminium nuclei in complex aluminium hydrides can also be obtained from static NMR spectra recorded using the Solomon echo pulse sequence.\cite{ZF24} 

\begin{figure}
\begin{center}
	\includegraphics[width=8cm]{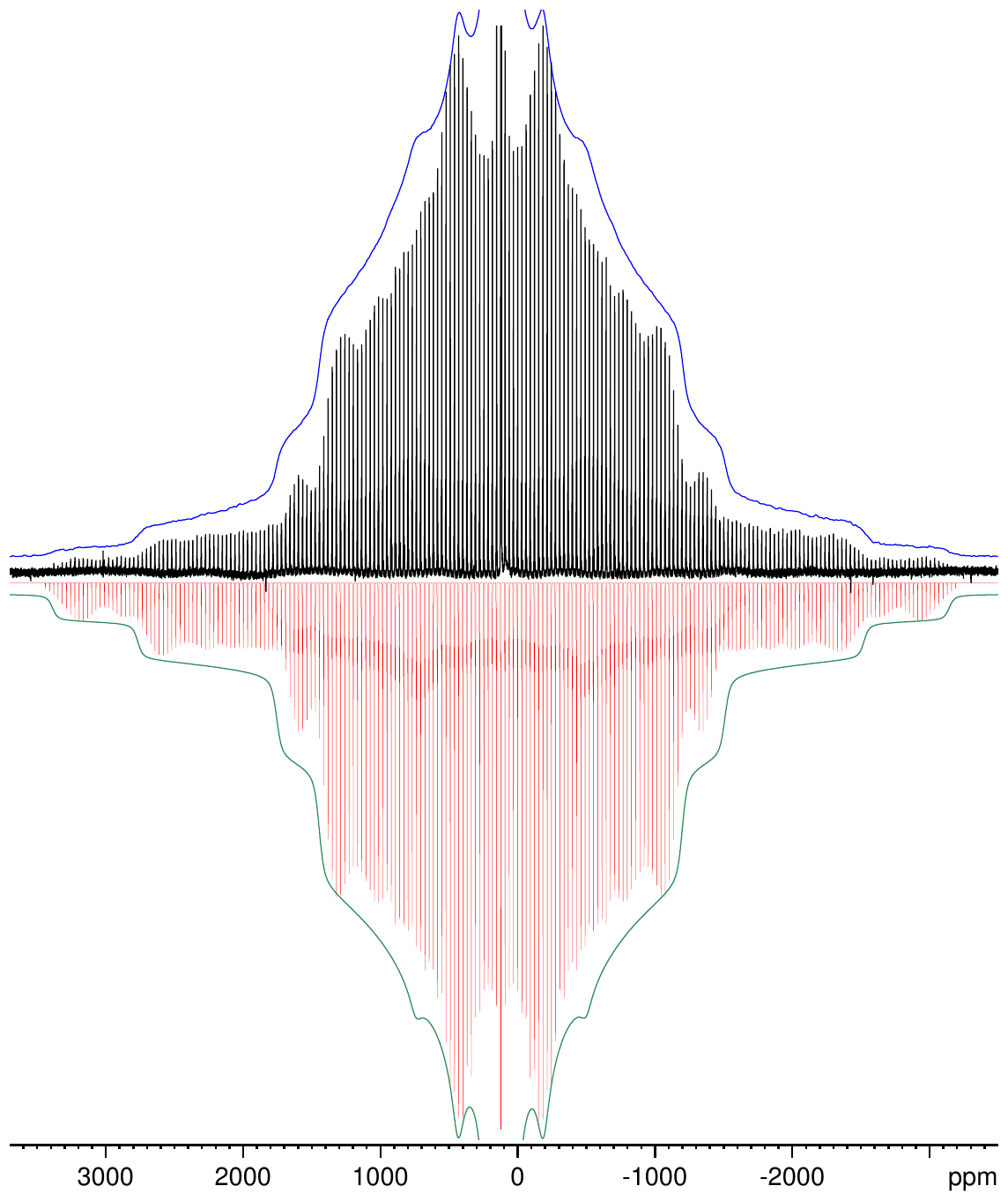}
  \caption{Experimental $^{27}$Al~NMR spectra of \ce{CsAlH4(\textit{o})} measured for a stationary sample (blue) and the same sample spinning at  $\nu_{\text{MAS}} = 4$\,kHz (black) and their simulations (green and red, respectively) with the parameters $\delta_{\text{iso}}=121.4$\,ppm, $C_{\text{Q}} = 1.42$\,MHz and $\eta = 0.62$. To ease the comparison, the simulated spectra have been inverted. A Lorentzian line broadening with $LB=5000$\,Hz and a Gaussian broadening with $GB=5000$ were used for the simulated static spectrum.}
  \label{fgr:CsAlH4_o}
\end{center}
\end{figure}

\begin{figure}
\begin{center}
	\includegraphics[width=8cm]{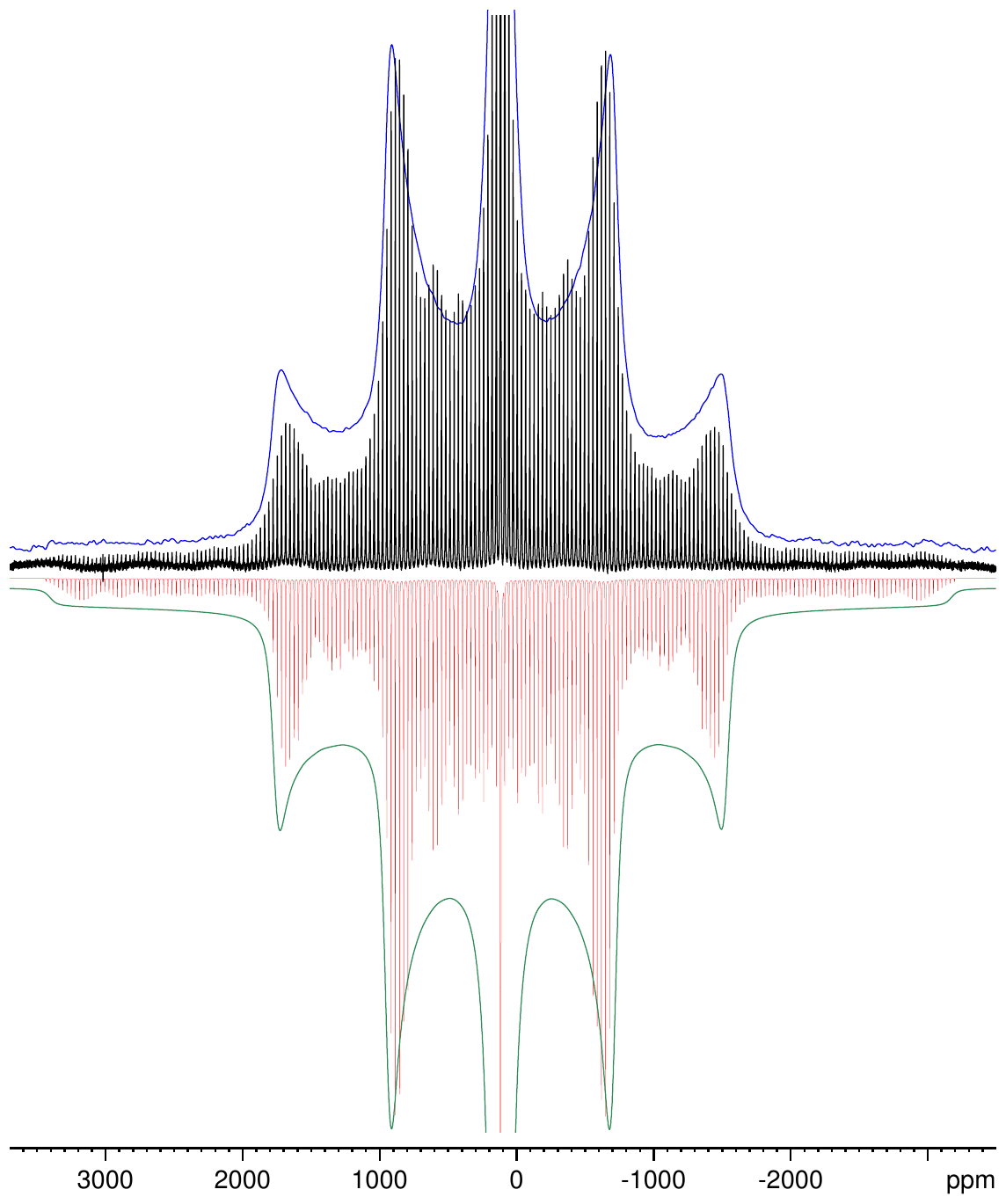}
  \caption{Experimental $^{27}$Al~NMR spectra of \ce{CsAlH4(\textit{t})} measured for a stationary sample (blue) and the same sample spinning at  $\nu_{\text{MAS}} = 4$\,kHz (black) and their simulations (green and red, respectively) with the parameters $\delta_{\text{iso}}=116.8$\,ppm, $C_{\text{Q}} = 1.42$\,MHz and $\eta = 0$. To ease the comparison, the simulated spectra have been inverted. A Lorentzian line broadening with $LB=8000$\,Hz and a Gaussian broadening with $GB=5000$ were used for the simulated static spectrum.}
  \label{fgr:CsAlH4_t}
\end{center}
\end{figure}

Figures~\ref{fgr:CsAlH4_o} and \ref{fgr:CsAlH4_t} show the experimental and simulated spectra using both approaches for the orthorhombic and tetragonal phase, respectively. The depicted lineshapes are typical of first-order quadrupole interaction. \cite{Man11} The positions of the characteristic discontinuities of the static lineshape as well as the intensity modulations of the spinning sidebands in the MAS spectra are obviously nicely reproduced by the simulations. The significantly reduced intensities in the outer wings of the experimental spectra are mainly caused by the insufficient excitation width of the finite pulses used. \cite{Freude00} Neither the finite excitation width nor the finite bandwidth of the NMR probe are taken into account by the simulation program used. The quality of the fits can be better judged from the enlarged versions of these figures provided as Figure~\,S1 and Figure~\,S2 in the supplementary material. The additional Figure~\,S3 shows that even subtle features in certain regions of the MAS NMR spectrum of \ce{CsAlH4(\textit{o})} are very well reproduced by the simulation.
The experimentally determined isotropic chemical shifts and quadrupole coupling parameters together with the estimated margins of error are summarized in Table \ref{tbl:EXP_vs_GIPAW}.

The values obtained for the quadrupole constant $C_{\text{Q}}$ by spectra simulation agree very well with the above estimate obtained from the spread of the spinning sidebands of the outer satellite transitions. The results corroborate that $C_{\text{Q}}$ is almost identical for the aluminium nuclei in both polymorphs. Since the local geometries around the aluminium nuclei are significantly different from each other, we regard this coincidence as an accidental one. On the other hand, the close resemblance of the data found for \ce{CsAlH4(\textit{o})} ($C_{\text{Q}}=1.42$\,MHz, $\eta=0.62$) with those found for \ce{KAlH4} ($C_{\text{Q}}=1.29$\,MHz, $\eta=0.64$)\cite{ZF19,ZF24} surely reflects the fact that these materials are isostructural.

As previously found for \ce{KAlH4} and \ce{NaAlH4},\cite{ZF24} we do not see any of the characteristic effects of a possible anisotropy of the chemical shift in the NMR spectra. The total spread of a resonance line due to the anisotropy of the chemical shift is given by the span $\varOmega$.\cite{Mason93} Our GIPAW calculations for the DFT-optimized geometries yield values of about 4.9\,ppm (see Table S4 in the supplementary material) and 2.6\,ppm (see Table S2) for the span, corresponding to about 640\,Hz and 340\,Hz, for \ce{CsAlH4(\textit{o})} and  \ce{CsAlH4(\textit{t})}, respectively.  These data would explain why the anisotropy of the chemical shift has no significant effect on the lineshapes, but contributes only to the line broadening of the static spectra.

\subsection{GIPAW vs Experiment}

The DFT-GIPAW calculated NMR parameters are included in Table \ref{tbl:EXP_vs_GIPAW}. While the  $\delta_{\text{iso}}$ value of \ce{CsAlH4(\textit{t})} is reproduced to within 0.5\,ppm, a larger deviation occurs for the orthorhombic phase, with $\delta_{\text{iso,DFT}}$ being about 2\,ppm smaller than the experimental (MAS\,NMR) value. Nevertheless, we consider this agreement satisfactory, especially as the calculations clearly reproduce the experimental observation that the two phases can be distinguished by virtue of the difference in $\delta_{\text{iso}}$.
A more complex picture emerges for the quadrupole coupling constant $C_{\text{Q}}$ and the asymmetry parameter $\eta$. The latter quantity equals 0 due to the local symmetry in the tetragonal phase, which agrees with the experiment. For CsAlH4(o), however, $\eta$ deviates significantly, being about 30\% smaller than the experimental value. The magnitude of $C_{\text{Q}}$ is consistently overestimated for both phases, with DFT-calculated values of +2.10\,MHz and $-2.05$\,MHz for \ce{CsAlH4(\textit{t})} and \ce{CsAlH4(\textit{o})}, respectively, compared to 1.42\,MHz measured experimentally (as pointed out above, the sign of $C_{\text{Q}}$ is not accessible from experiment). A very similar overestimation of $C_{\text{Q}}$ was observed for orthorhombic \ce{KAlH4} in our previous work.\cite{ZF24}

\begin{table}
  \caption{Comparison of the results of DFT-GIPAW calculations  with the experimental data from NMR spectra measured with and without MAS using either single-pulse excitation (SPE) or the Solomon echo pulse sequence (SE).}
  \label{tbl:EXP_vs_GIPAW}
  \begin{tabular}{llccc}
    \hline
    Material  & Method &  $\delta_{\text{iso}}$ [ppm] & $C_{\text{Q}}$ [MHz] & $\eta$\\
    \hline
    \ce{CsAlH4(\textit{o})}  & SPE, MAS & $121.4\pm0.2$ & $1.42\pm0.02$ &  $0.60\pm0.02$  \\
     &  SE, static & $122\pm5$ & $1.42\pm0.01$ & $0.62\pm0.01$\\
     &  GIPAW & 119.2 & $-2.05$ & 0.44\\
    \ce{CsAlH4(\textit{t})}  & SPE, MAS &  $116.8\pm0.2$ & $1.42\pm0.02$ &  $<$0.05\\
     & SE, static & $116\pm5$ & $1.43\pm0.02$ & $<$0.03\\
     & GIPAW & 116.2 & 2.10 & 0\\
    \hline
  \end{tabular}
\end{table}

Regarding the observed deviations in $C_{\text{Q}}$ and, for \ce{CsAlH4(\textit{o})}, $\eta$ between DFT-GIPAW calculations and experiment, one  has to consider that the DFT-calculated NMR parameters were obtained for a statically optimized structure at 0 K. In the NMR experiments, which were performed at room temperature, thermal motion of the atoms will result in changes of the average atomic positions with respect to absolute zero, in turn affecting the observed NMR parameters. In order to gauge the impact of changes in the local structure of the \ce{AlH4} tetrahedra on the DFT-GIPAW results, a series of calculations was performed for structures in which either the Al--H bond distance or the H--Al--H angles were varied systematically. 

\begin{figure}
\begin{center}
	\includegraphics[width=16cm]{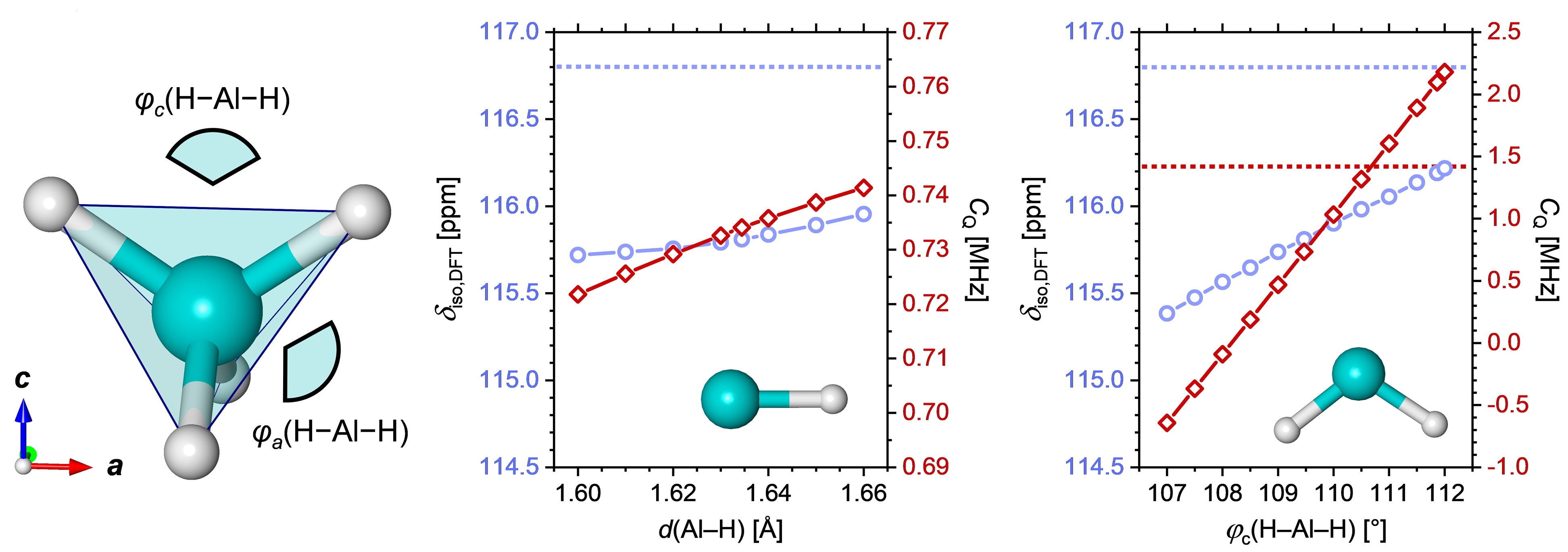}
  \caption{Left: \ce{AlH4} tetrahedron in \ce{CsAlH4(\textit{t})}. Center and right: DFT-GIPAW values of $\delta_{\text{iso,DFT}}$ and $C_{\text{Q}}$ computed for \ce{CsAlH4(\textit{t})} structures in which the Al--H distance (center) or the H--Al--H angles (right) were varied. The horizontal dashed lines indicate the experimentally determined values.}
  \label{fgr:DFT_tetra}
\end{center}
\end{figure}

For the tetragonal phase, all four Al--H bonds are equivalent, and there are only two non-equivalent angles $\varphi_{\text{c}}\text{(H--Al--H)}$ and $\varphi_{\text{a}}\text{(H--Al--H)}$ (see left panel of Figure~\ref{fgr:DFT_tetra}). For this structure, the Al--H bond distance was varied from 1.60\,\AA\ to 1.66\,\AA, (DFT-optimized value: 1.6344 \AA), fixing all tetrahedral angles to the ideal value of 109.47$^{\circ}$. The results, visualized in the central panel of Figure~\ref{fgr:DFT_tetra}, show that neither $\delta_{\text{iso,DFT}}$ nor $C_{\text{Q}}$ are particularly sensitive to the Al--H distance, with maximal changes in the order of 0.25\,ppm and 0.02\,MHz, respectively. In further calculations for this phase, the distance was fixed to the DFT-optimized value, and the angle $\varphi_{\text{c}}\text{(H--Al--H)}$ was varied from 107 to 112$^{\circ}$ (DFT-optimized value: 111.86$^{\circ}$, note that $\varphi_{\text{a}}\text{(H--Al--H)}$ is not independent and varies accordingly). The results, shown in the right panel of Figure~\ref{fgr:DFT_tetra}, reveal a prominent linear dependence of $C_{\text{Q}}$ on the angle $\varphi_{\text{c}}\text{(H--Al--H)}$, with $C_{\text{Q}}$ increasing by about 0.57\,MHz per degree. By interpolation, one can infer that perfect agreement with experiment would be reached at a $\varphi_{\text{c}}\text{(H--Al--H)}$ angle of about 110.7$^{\circ}$. 

The influence of changes in the Al--H distance and the angle $\varphi_{\text{c}}\text{(H--Al--H)}$ on the chemical shift anisotropy are illustrated in Figure S4 in the supplementary material. A variation of $d$(Al--H) results in only insignificant changes of the span $\varOmega$, but its dependence on the $\varphi_{\text{c}}\text{(H--Al--H)}$ angle is much more pronounced: While $\sigma_{xx}$ decreases with increasing angle, $\sigma_{zz}$ increases, resulting in a decrease of $\varOmega$ from 20.4\,ppm to 2.1\,ppm across the range of angles considered. Since the DFT-optimized value falls close to the upper boundary value, the span computed for this geometry is small, amounting to 2.6\,ppm. For the above mentioned value of $\varphi_{\text{c}}\text{(H--Al--H)}$ (110.7$^{\circ}$) that would result in a perfect agreement between the experimental and the calculated quadrupole coupling constant, a span of about 6 ppm is obtained. The full results of the DFT-GIPAW calculations for these \ce{CsAlH4(\textit{t})} structures are compiled in Tables S1 and S2 in the supplementary material.

\begin{figure}
\begin{center}
	\includegraphics[width=16cm]{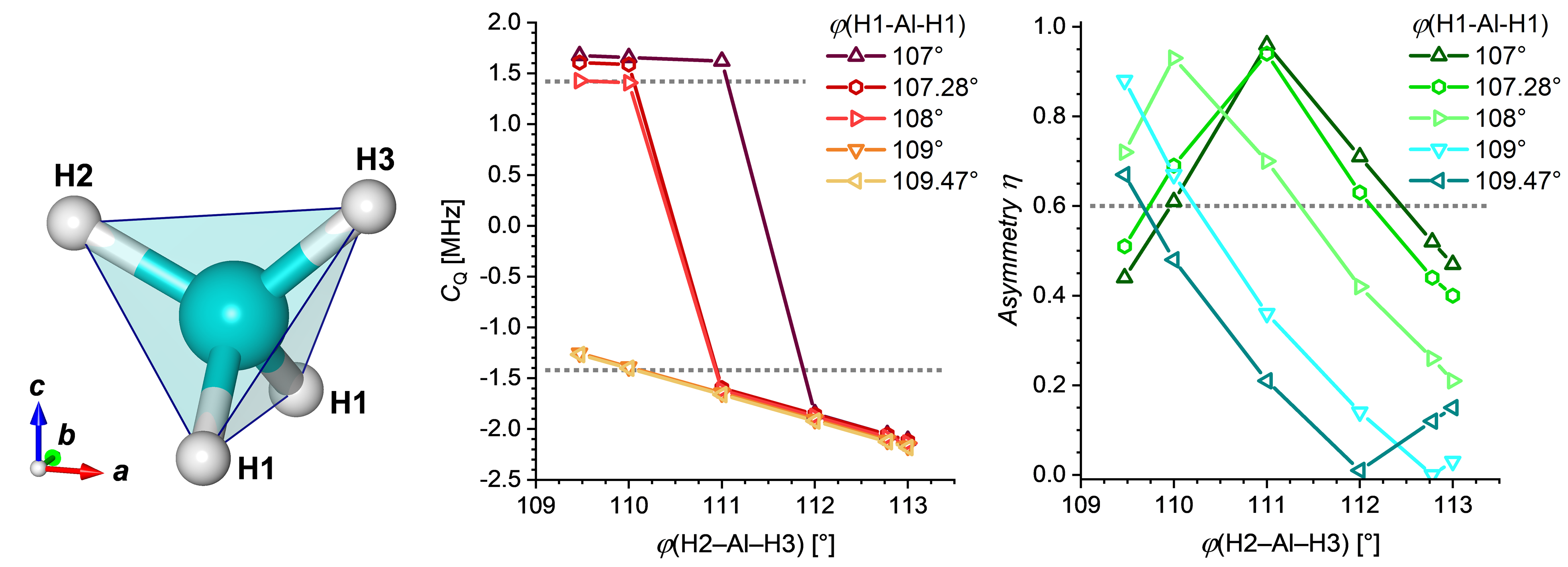}
  \caption{Left: \ce{AlH4} tetrahedron in \ce{CsAlH4(\textit{o})}. Center and right: DFT-GIPAW values of $C_{\text{Q}}$ (center) and $\eta$ (right) computed for \ce{CsAlH4(\textit{o})} structures in which the H--Al--H angles were varied. }
  \label{fgr:DFT_ortho}
\end{center}
\end{figure}

For \ce{CsAlH4(\textit{o})}, the reduced local symmetry, with three non-equivalent H atoms (Figure~\ref{fgr:DFT_ortho}), renders a comprehensive sampling of the possible local geometries more challenging. Since the calculations for \ce{CsAlH4(\textit{t})} indicated only a small influence of the Al--H distance, the bond distances were fixed to the DFT-optimized values ($d$ = 1.6402/1.6338/1.6309\,\AA\ for Al--H1/H2/H3). The two angles $\varphi\text{(H1--Al--H1)}$ and $\varphi\text{(H2--Al--H3)}$ were varied from 107$^{\circ}$ to 109.47$^{\circ}$ (DFT-optimized value: 107.28$^{\circ}$) and from 109.47$^{\circ}$ to 113$^{\circ}$ (DFT-optimized value: 112.78$^{\circ}$), respectively. The full results of the DFT-GIPAW calculations for these \ce{CsAlH4(\textit{o})} structures are compiled in Tables S3, S4 and S5 in the supplementary material. Again, we note only a minimal influence of changes in the angles on the isotropic chemical shift $\delta_{\text{iso,DFT}}$, which varies between 119.2\,ppm and 119.4\,ppm. As in the case of \ce{CsAlH4(\textit{t})}, the variation of the chemical shift anisotropy is more pronounced, with $\varOmega$ ranging from 4.3\,ppm to 12.9\,ppm across the range of angles considered.

As is visible in the central and right panels of Figure~\ref{fgr:DFT_ortho}, the magnitude of the H--Al--H angles has a significant impact on $C_{\text{Q}}$ and $\eta$. The abrupt change in the sign of $C_{\text{Q}}$ that is observed with continuous change of $\varphi\text{(H2--Al--H3)}$ does not reflect a sudden change in a physical property, but is a direct consequence of the suboptimal definition of $C_{\text{Q}}$,
\begin{equation}
  C_{\text{Q}}=\frac{e^2qQ}{h},
   \label{eq:CQ}
\end{equation}
where $eQ$ is the electric quadrupole moment of the nucleus and $eq=V_{zz}$ the $zz$-component of the EFG tensor at the site of the nucleus. To ensure that the asymmetry parameter $\eta$, 
\begin{equation}
  \eta=\frac{V_{xx}-V_{yy}}{V_{zz}},
   \label{eq:eta}
\end{equation}
is in the range $0\leq\eta\leq1$, the labels of the axes in the principal axes system of the EFG tensor are chosen so that the condition $\lvert V_{zz}\rvert\geq\lvert V_{yy}\rvert\geq\lvert V_{xx}\rvert$ is satisfied.\cite{FreudeHaase,quadnmr} With the definition of the quadrupole coupling  $C_{\text{Q}}$ as being proportional to the EFG eigenvalue of maximum magnitude (equation \ref{eq:CQ}), a small change in the geometry is likely to change the sign of $C_{\text{Q}}$ if $\eta$ is close to 1. This peculiarity is the reason why Figure~\ref{fgr:DFT_ortho} may be a bit confusing, at least at first sight. The evolution of the tensor components is not apparent. A  representation of the quadrupole coupling constant that removes the discontinuities could be achieved by plotting  
 $|{C_{\text{Q}}|}$ instead of $C_{\text{Q}}$. However, we want to propose an alternative approach near the end of this section.

The asymmetry parameter $\eta$ exhibits a very pronounced sensitivity towards the H--Al--H angles, in some cases changing by 0.25 and more when one of the angles changes by one degree. When attempting to match the experimentally measured values of $C_{\text{Q}}$ and $\eta$, a structure with $\varphi\text{(H1--Al--H1)}$ = 109$^{\circ}$ and $\varphi\text{(H2--Al--H3)}$ = 110$^{\circ}$ would deliver good agreement, with $C_{\text{Q}} = -1.39$\,MHz and $\eta = 0.67$ (Table S3). For this geometry a span $\varOmega$ of about 10 ppm is obtained (Table S4), which would be still too small to cause significant effects on the lineshape of the satellite transitions. However, we note that only two of the angles were varied systematically, and that the remaining angles of the \ce{AlH4} tetrahedron depend on them. With this in mind, it seems reasonable to expect that other local geometries which remained untested here could deliver $C_{\text{Q}}$ and $\eta$ values that are in similarly good agreement with experiment. 

\begin{figure}
\begin{center}
	\includegraphics[width=8cm]{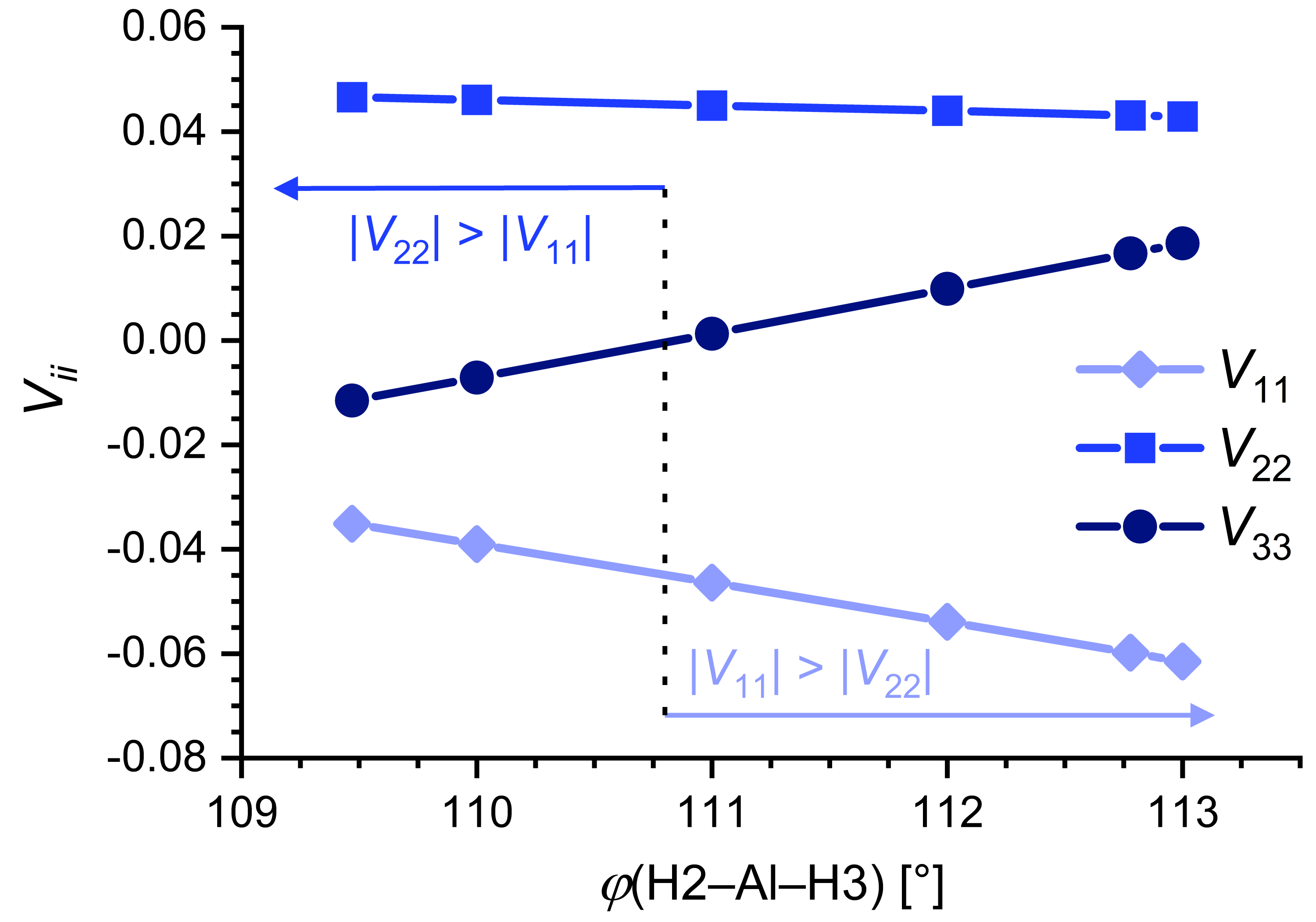}
  \caption{Evolution of the eigenvalues $V_{11}$, $V_{22}$, and $V_{33}$ of the EFG tensor of \ce{CsAlH4(\textit{o})} as a function of the $\varphi\text{(H2--Al--H3)}$ angle. The shown values were obtained for a constant $\varphi\text{(H1--Al--H1)}$ angle of 107.28$^{\circ}$. The tensor components are given in  a.u.\,(atomic unit of electric field gradient: $0.97173\cdot10^{22}\,$Vm$^{-2}$).}
 
  \label{fgr:Vii_angle}
\end{center}
\end{figure}

Figure~\ref{fgr:Vii_angle} plots the eigenvalues of the EFG tensor of \ce{CsAlH4(\textit{o})}, using $V_{ii}$ with $i$ = 1, 2, 3 as labels for the eigenvalues associated with the eigenvectors found in the diagonalisation. These eigenvectors are roughly, but not in all cases exactly, parallel to the unit cell axes $a$, $b$, and $c$. 
While only the results for one value of the $\varphi\text{(H1--Al--H1)}$ angle are shown in that figure, the full results are provided in Table S5 in the supplementary material, where also a more precise description of the eigenvectors can be found. 

The $V_{ii}$ shown in Figure~\ref{fgr:Vii_angle} vary continuously with $\varphi\text{(H2--Al--H3)}$. For the two smallest angles, $V_{22}$ is the component having the largest absolute value, resulting in positive values of $V_{zz}$. Since the quadrupole moment of aluminium ($Q(^{27}\text{Al})=146.6\,$mbarn\cite{Pyykko18}) is positive, the quadrupole coupling constant  $C_{\text{Q}}$ has the same sign as $V_{zz}$. Hence, $C_{\text{Q}}$ is positive in this range of $\varphi\text{(H2--Al--H3)}$. Above an angle of about 110.8$^{\circ}$ (estimated by interpolation of the $V_{ii}$ values), the absolute value of $V_{11}$ is the largest and $V_{11}$ becomes $V_{zz}$ according to the labelling convention given above.  Consequently,  $C_{\text{Q}}$ is also negative at these larger $\varphi\text{(H2--Al--H3)}$ angles.

Figure~\ref{fgr:Vii_angle} shows that the maximum difference of the EFG eigenvalues seems to be a useful parameter for the comparison of calculated and experimental data. In analogy to the definition of the span $\varOmega$ and skew $\kappa$ to describe the anisotropic part of the chemical shift,\citep{Mason93} one could, for instance, define a parameter $\varOmega_{\text{Q}}$ and a parameter $\kappa_{\text{Q}}$ with the following equations:
\begin{equation}
  \varOmega_{\text{Q}}=\frac{eQ}{h} (V_{zz}-V_{xx})
   \label{eq:OmegaQ}
\end{equation}
and
\begin{equation}
  \kappa_{\text{Q}}=\frac{-3V_{yy}}{V_{zz}-V_{xx}}.
   \label{eq:kappaQ}
\end{equation}
Here, the labels of the axes in the principal axes system of the EFG tensor should be chosen so that the condition  $V_{zz} \geq V_{yy} \geq V_{xx}$ is satisfied. The parameter $\varOmega_{\text{Q}}$ is now proportional to the maximum difference of the EFG eigenvalue and should vary continuously with small changes in the geometry.  $\kappa_{\text{Q}}$ varies from $-1$ ($V_{yy} = V_{zz}$) over 0 ($V_{xx} = -V_{zz}$) to 1 ($V_{xx} = V_{yy}$).

\begin{figure}
\begin{center}
	\includegraphics[width=16cm]{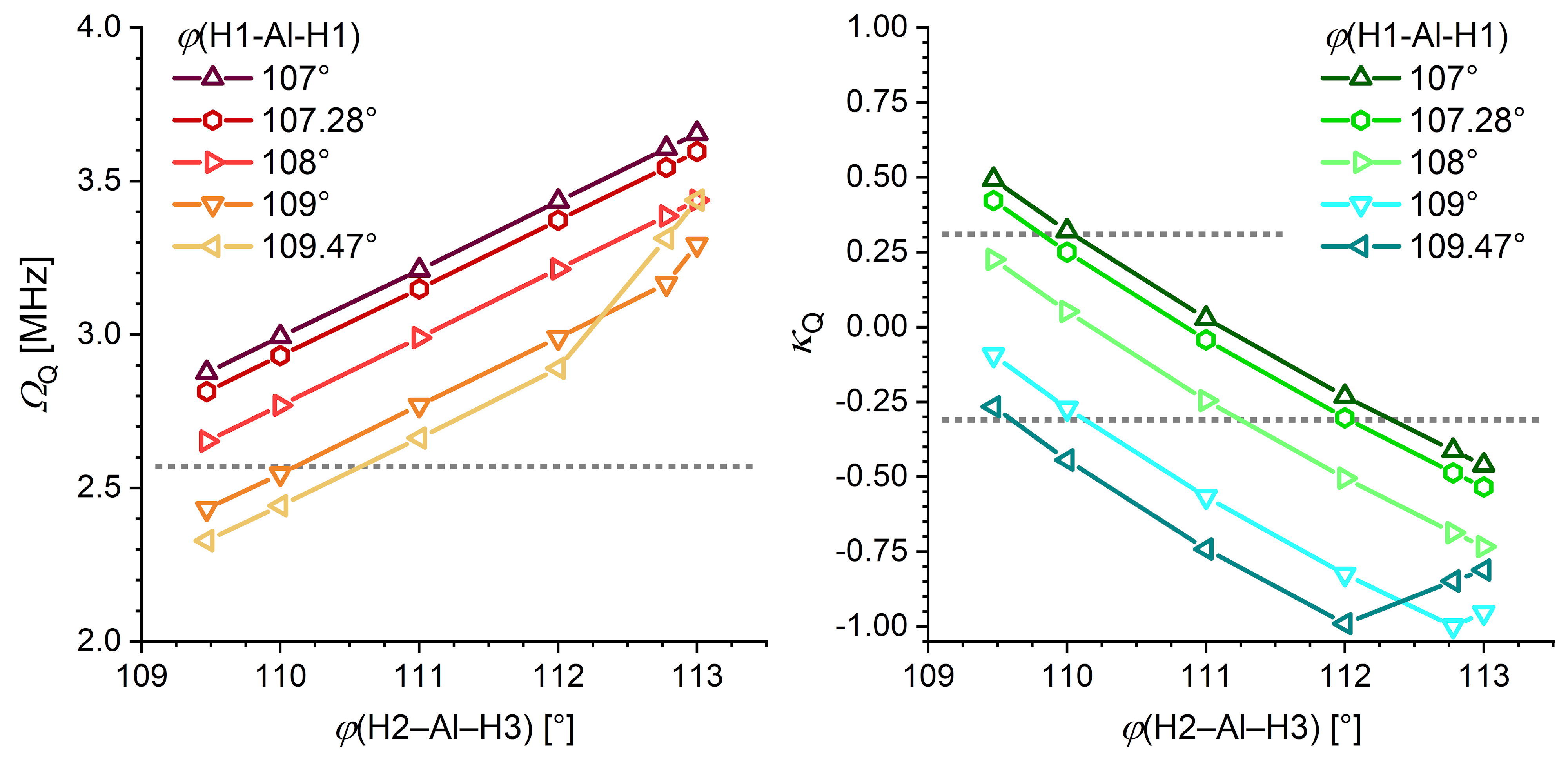}
  \caption{DFT-GIPAW values of $\varOmega_{\text{Q}}$ (left) and $\kappa_{\text{Q}}$ (right) computed for \ce{CsAlH4(\textit{o})} structures in which the H--Al--H angles were varied.}
 
  \label{fgr:Omega&kappa_Q}
\end{center}
\end{figure}

Figure~\ref{fgr:Omega&kappa_Q} plots these newly defined parameters for \ce{CsAlH4(\textit{o})} that were obtained for the same combinations of $\varphi\text{(H1--Al--H1)}$ and $\varphi\text{(H2--Al--H3)}$ as used in Figure~\ref{fgr:DFT_ortho}. The experimentally determined values for $C_{\text{Q}}$ and $\eta$ (Table~\ref{tbl:EXP_vs_GIPAW}) correspond to $\varOmega_{\text{Q}}=2.57$\,MHz and $|{\kappa_{\text{Q}}}|= 0.31$. In line with the above finding, a rather good agreement between NMR experiment and calculation can obviously be achieved with $\varphi\text{(H1--Al--H1)} \approx 109^{\circ}$ and $\varphi\text{(H2--Al--H3)} \approx 110^{\circ}$. The actual bond angles $\varphi\text{(H--Al--H)}$ are apparently closer to the bond angle in a perfect tetrahedral molecule than those found by DFT optimization.

In our previous study of orthorhombic \ce{KAlH4}, we obtained experimental values of $C_{\text{Q}} = 1.30$\,MHz and $\eta = 0.64$, compared to DFT-GIPAW results of $C_{\text{Q}} = -1.69$\,MHz and $\eta = 0.59$.\cite{ZF24} It seems like a logical extension of the above investigation to test whether a variation of the H--Al--H angles for this compound brings about similarly large changes of the quadrupole coupling constant and the asymmetry as for \ce{CsAlH4(\textit{o})}. The results of a set of calculations varying the $\varphi\text{(H1--Al--H1)}$ and $\varphi\text{(H2--Al--H3)}$ angles are presented in Figure S6 and Table S6 in the supplementary material, which show that both quantities vary significantly across the range of angles studied. Again, relatively modest changes in both angles away from their DFT-optimized values would be sufficient to simultaneously match the experimental values of $C_{\text{Q}}$ and $\eta$.

Finally, we take a brief look at the \textsuperscript{133}Cs NMR chemical shifts obtained from the DFT-GIPAW calculations. The calculated $\delta_{\text{iso,DFT}}$ values amount to 108.0 and 165.3\,ppm  for \ce{CsAlH4(\textit{t})} and \ce{CsAlH4(\textit{o})} ($\Delta(\delta_{\text{iso,DFT}})=57.3$ ppm), with the corresponding experimental values being 133.7\,ppm and 180.7\,ppm  ($\Delta(\delta_{\text{iso}})=47.0$\,ppm).\cite{Krech14} In the view of the simple relationship used to convert calculated chemical shieldings into chemical shifts (equation \ref{eq:delta_Cs}), the agreement between DFT and experiment appears acceptable. We note in this context that a recent study of \textsuperscript{133}Cs NMR chemical shifts of clay minerals \cite{Ohkubo23} proposed an equation with a slope $\alpha$ smaller than unity  for this conversion. Interestingly, a multiplication of the $\Delta(\delta_{\text{iso,DFT}})$ value obtained in the present work by the $\alpha$ value of 0.832 proposed in that earlier study results in a $\Delta(\delta_{\text{iso,DFT}})$ value of 47.7\,ppm, in near-perfect agreement with experiment. A more sophisticated treatment would require the use of a reference set consisting of several structures to obtain a refined relationship, as done for the \textsuperscript{27}Al NMR chemical shifts in our prior work.\cite{ZF24} However, the extension of this approach to Cs-containing compounds goes beyond the scope of the present study.

\section{Conclusions}
In line with our earlier paper on complex aluminium hydrides,\cite{ZF24} we have shown that the parameters of the quadrupole coupling can be determined from \textsuperscript{27}Al NMR spectra measured for stationary samples with at least the same precision as from MAS~NMR spectra. Good estimates of these parameters, directly read off the experimental spectra, 
can be used as starting values for spectra simulation. For the determination of the isotropic chemical shift, the MAS approach is superior. However, avoiding fast sample spinning during the NMR measurement would be of paramount importance when less stable materials \cite{Cao24} are to be investigated that could decay under mechanical stress and/or temperature increase due to frictional heating. 

The structural differences between the two polymorphs investigated are clearly mirrored in the spectra of the satellite transitions. Both the lineshapes in the static spectra and the spinning sideband patterns in the MAS spectra are strikingly different, making their identification for samples containing both polymorphs possible. In accordance with the experimental result, almost identical magnitudes for the quadrupole coupling constant $C_{\text{Q}}$ of both polymorphs were obtained by GIPAW calculations using DFT-optimized structures. However, the magnitude of $C_{\text{Q}}$ was overestimated by about 45\%. 

A systematic variation of the Al--H distances and H--Al--H angles showed that $C_{\text{Q}}$ is hardly affected by the bond length, whereas even relatively minor variations in the bond angles result in large variations of $C_{\text{Q}}$ and $\eta$. While the isotropic chemical shift 
$\delta_{\text{iso}}$ of the aluminium nuclei is not sensitive to the H--Al--H angles, the chemical shift anisotropy shows a more pronounced dependence.  However, it might remain too small to be observed experimentally at the magnetic field strengths that are usually applied.
	
	We have to note that the DFT-optimized structures correspond to the energy minima at 0\,K. Thermal motion will lead to changes in the average structure on the timescale that is probed by NMR experiments. We expect that thermal motion will lead to a decrease of the distortion of the \ce{AlH4} tetrahedra compared to the structures calculated for 0\,K. This is supported by the observation that experiment and DFT nicely agree for geometries in which the H--Al--H angles are closer to the ideal value than in the DFT-optimized structures.

 The present work employed a ``trial-and-error'' approach based on a systematic variation of the angles in the DFT calculation. While this approach is very helpful to analyze existing experimental data, its predictive capabilities are severely limited. Predictions could be achieved by using average structures from DFT-based \textit{ab initio} molecular dynamics (AIMD) or by computing average values over NMR parameters computed for a sufficiently large ensemble of AIMD snapshots; however, any AIMD-based approach increases the computational demand by orders of magnitude. \cite{Vanlommel22} Even when such an approach is followed, the strong dependence of $C_{\text{Q}}$ on the geometry of the AlH\textsubscript{4} tetrahedra means that slight deviations between the computed average structure and the ``true'' structure could lead to non-negligible discrepancies between DFT and experiment, rendering accurate predictions for such compounds extremely challenging. 
It should be noted that the direct use of experimental structure data for the calculation of NMR parameters  without DFT optimization of the atomic coordinates is no promising alternative, in particular when hydrogen atoms are involved. In accordance with literature,\cite{Bonhomme2012, Ashbrook2016} we found in our previous paper \cite{ZF24} on complex aluminium hydrides that this approach results in much bigger deviations from the experimental quadrupole coupling parameters.

When the influence of systematic changes in the geometry on the quadrupole coupling are studied, the use of the parameter $C_{\text{Q}}$ and $\eta$ has the drawback that in the region of $\eta \approx 1$ rather small changes in the geometry can result in a change of the sign of $C_{\text{Q}}$. This discontinuity could be avoided when $|{C_{\text{Q}}|}$ instead of $C_{\text{Q}}$ is discussed. However, more appropriate seems to be the use of parameters that are defined along the lines of those used for the description of the chemical shift anisotropy: the span $\varOmega$ and the skew $\kappa$.\cite{Mason93} 
Of course, the sign of the newly defined skew parameter $\kappa_{\text{Q}}$ is not accessible by the NMR experiments described here. It is interesting to note that already in the paper by Mason,\cite{Mason93} the parameters span and skew were not restricted to the field of chemical shift anisotropy, but referred to as ``useful properties also for other tensor quantities contributing to the nuclear spin Hamiltonian''. Alternatively, one could discuss the diagonal elements of the EFG tensor. However, since the $V_{ii}$ obey the Laplace equation, there are only two independent parameters.

\subsection{Conflict of interest}
There are no conflicts to declare.

\begin{acknowledgement}

The authors thank Dr Daniel Krech, then at the MPI f\"ur Kohlenforschung in M\"ulheim an der Ruhr, for the preparation of the aluminium hydride samples used and Prof.\ Dieter Freude (Leipzig) for a stimulating discussion on the comparison of the experimental and calculated data obtained. MF gratefully acknowledges funding by the Deutsche Forschungsgemeinschaft (German Research Foundation, DFG) through a Heisenberg fellowship (project no. 455871835). 

\end{acknowledgement}

\subsection{Supplementary Material}
Supplementary material is available. 

\subsection{ORCID}
Bodo Zibrowius: 0000-0002-3894-7318\\
Michael Fischer: 0000-0001-5133-1537

\newpage
\bibliography{Poly_CsAlH4}

\end{document}